\documentclass[twocolumn,aps,prl,preprintnumbers,superscriptaddress]{revtex4}
\usepackage{}
\usepackage{bbm}
\usepackage[latin9]{inputenc}
\usepackage{amsmath}
\usepackage{amssymb}
\usepackage{graphicx}
\usepackage{mathrsfs}
\usepackage{amsfonts}
\usepackage{amsthm}
\usepackage{color}
\usepackage{txfonts}
\usepackage{lipsum}
\usepackage{ulem}
\usepackage{gensymb}
\usepackage{booktabs}
\usepackage[colorlinks=true,citecolor=blue,linkcolor=blue,urlcolor=blue,anchorcolor=blue]{hyperref}%
\hypersetup{colorlinks=true,citecolor=blue,linkcolor=blue,urlcolor=blue}
\providecommand{\U}[1]{\protect\rule{.1in}{.1in}}
\makeatletter

\newcommand{\Rmnum}[1]{\expandafter\@slowromancap\romannumeral #1@}
\makeatother
\makeatletter
\@ifundefined{textcolor}{}
{
\definecolor{BLACK}{gray}{0}
\definecolor{WHITE}{gray}{1}
\definecolor{RED}{rgb}{1,0,0}
\definecolor{GREEN}{rgb}{0,1,0}
\definecolor{BLUE}{rgb}{0,0,1}
\definecolor{CYAN}{cmyk}{1,0,0,0}
\definecolor{MAGENTA}{cmyk}{0,1,0,0}
\definecolor{YELLOW}{cmyk}{0,0,1,0}
}
\makeatother
\begin{document}
\title{Low-lying magnon frequency comb in skymion crystals}
\author{Xuejuan Liu}
\affiliation{School of Physics and State Key Laboratory of Electronic Thin Films and Integrated Devices, University of Electronic Science and Technology of China, Chengdu 610054, China}
\affiliation{School of Healthcare Technology, Chengdu Neusoft University, Chengdu 611844, China}
\affiliation{College of Physics and Engineering, Chengdu Normal University, Chengdu 611130, China}
\author{Zhejunyu Jin}
\affiliation{School of Physics and State Key Laboratory of Electronic Thin Films and Integrated Devices, University of Electronic Science and Technology of China, Chengdu 610054, China}
\author{Zhengyi Li}
\affiliation{School of Physics and State Key Laboratory of Electronic Thin Films and Integrated Devices, University of Electronic Science and Technology of China, Chengdu 610054, China}
\author{Zhaozhuo Zeng}
\affiliation{School of Physics and State Key Laboratory of Electronic Thin Films and Integrated Devices, University of Electronic Science and Technology of China, Chengdu 610054, China}
\author{Minghao Li}
\affiliation{School of Physics and State Key Laboratory of Electronic Thin Films and Integrated Devices, University of Electronic Science and Technology of China, Chengdu 610054, China}
\author{Yuping Yao}
\affiliation{School of Physics and State Key Laboratory of Electronic Thin Films and Integrated Devices, University of Electronic Science and Technology of China, Chengdu 610054, China}
\author{Yunshan Cao}
\affiliation{School of Physics and State Key Laboratory of Electronic Thin Films and Integrated Devices, University of Electronic Science and Technology of China, Chengdu 610054, China}
\author{Yinghui Zhang}
\affiliation{School of Healthcare Technology, Chengdu Neusoft University, Chengdu 611844, China}
\author{Peng Yan}
\email[Corresponding author: ]{yan@uestc.edu.cn}
\affiliation{School of Physics and State Key Laboratory of Electronic Thin Films and Integrated Devices, University of Electronic Science and Technology of China, Chengdu 610054, China}
\begin{abstract}
A stable, low-power and tunable magnon frequency comb (MFC) is crucial for magnon-based precision measurements, quantum information processing and chip integration. Original method for creating MFC utilizes the nonlinear interactions between propagating spin waves and localized oscillations of an isolated magnetic texture, e.g., skyrmion. It requires a driving frequency well above the ferromagnetic resonance (FMR) and the spectrum frequency of MFC will quickly approach to the detection limit of conventional microwave technique after only tens of  comb teeth. In addition, the detection and manipulation of a single skyrmion is challenging in experiments due to its high degree of locality. These issues hinder the applications of MFC. In this work, we report the low-lying MFC with comb frequencies below the FMR in a skyrmion crystal (SkX). We show that the MFC originates from the three-wave mixing between the collective skyrmion gyration and breathing in the SkX. Our findings significantly improve the efficiency of the nonlinear frequency conversion from a single-frequency mircowave input, and establish a synergistic relationship between the SkX and MFC, which paves the way to coherent information processing and ultra-sensitive metrology based on MFC.
\end{abstract}

\maketitle
\section{INTRODUCTION}
Optical frequency comb (OFC) is a wide spectrum composed of a series of discrete, equally spaced and phase-locked lasers in the frequency domain \cite{Udem,DelHaye}. Due to its excellent time-frequency reference characteristics, highly coherent and stable output, OFC is widely used in important fields such as atomic clocks \cite{Ludlow2015}, satellite navigation \cite{Lezius2016}, low-noise microwave source and optical spectroscopy \cite{Picque2019}. The widespread applications of OFCs have greatly inspired researchers to explore other types of frequency combs \cite{Wu2022,Ganesan2017}. In the realm of magnetism, researchers have proposed various techniques for generating magnon frequency combs (MFCs) \cite{Hula2022,Liu2023,zhou2021,Rao2023,
Wang2021,Wang2022,JZJY2023PRL,Yao2023,Xu2023PRL,
WangNP2024,LiuPRB2024,Shennpj2024}. Most studies of the MFC, however, focused on the three-magnon or four-magnon interaction between a progagating magnon and localized mode associated with a single skyrmion, vortex, or bimeron \cite{Wang2021,Wang2022,JZJY2023PRL,Yao2023,
LiuPRB2024,Shennpj2024,WangJ2022,Liang2024,ZhangAPL2024}. The frequencies of MFC are higher than the ferromagnetic resonance (FMR), and may soon approach to the detection limit of conventional microwave technique after tens of comb teeth. In addition, a single skyrmion is challenging to detect and manipulate in experiments. These issues hinder the realization of an efficient, on-chip integrated, and tunable MFC-based magnonic functional device. A low-lying MFC is thus highly demanded.

Compared to a single skyrmion, the skyrmion crystal (SkX) is more common and stable \cite{Yu2010,Heinze2011,Yamasaki2015,Louise2023,Matsuki2023,Okuyama2024}. Importantly, both the gyration and breathing frequencies of SkX fall far below the FMR. Because of the orthogonality between the breathing and gyration modes and their distinguishing frequencies, it is usually believed that they cannot couple with each other via a resonant way \cite{LiuAIP2020,Chengprb2024}. One question naturally arises: Is it possible to couple the skyrmion breathing and gyration modes in a nonlinear way and to form the MFC in SkX?

In this work, we give a firm answer to the above question. We develop a theoretical framework to describe the collective dynamics of SkX and apply it to model the dispersion of the gyration and breathing modes and their nonlinear three-wave mixing. The derived band structure is verified by full micromagnetic simulations with excellent agreement. To observe the low-lying MFC in SkX, we apply a \textit{single-frequency} microwave field along the out-of-plane direction to first excite the skyrmion breathing. In SkX with a confined geometry, the collective breathing motion of skyrmions forms a standing wave due to the skyrmion-skyrmion coupling. It is found that the skyrmion does not breathe but gyrate at the nodes of the standing wave. The standing-wave skyrmion breathing and the skyrmion gyration at the nodes then strongly mix with each other and finally produce the MFC. The frequency spacing of MFC is equal to the gyration frequency of the node skyrmion. We observe this process both in one-dimensional (1D) and two-dimensional (2D) SkXs, which demonstrates an universal mechanism to generate the low-lying MFC in SkXs.

The paper is organized as follows: In Sec. \ref{theoretical model}, we develop the theoretical framework to describe both the linear magnon band structure and nonlinear interaction between the breathing and gyrating modes in the SkX. We implement micromagnetic simulations to verify our theoretical predictions in Sec. \ref{Micromagnetic simulations}. In Sec. \ref{conclusion}, we conclude this work.

\section{phenomenological theory} \label{theoretical model}

The magnetization dynamics is governed by the phenomenological Landau-Lifshitz Gilbert (LLG) equation
 \begin{equation}\label{Eq1}
    \frac{\partial \mathbf{m}}{\partial t}=-\gamma\mathbf{m}\times {\mathbf{H}_{\texttt{eff}}}+\alpha\mathbf{m}\times\frac{\partial\mathbf{m}}{\partial{t}},
\end{equation}where $\mathbf{m}$ is the unit magnetization vector, $t$ is the time, $\gamma$ is the gyromagnetic ratio, and $\alpha$ is for the phenomenological Gilbert damping constant. The effective field $\mathbf{H}_{\texttt{eff}}$ is the functional derivative of the total magnetic energy $W$, $\mathbf{H}_{\texttt{eff}}=-\frac{1}{\mu_{0}M_{s}} \frac{\delta W}{\delta \mathbf{m}}$, where $M_{s}$ is the saturation magnetization, $W$ consists of the exchange energy $E_{\texttt{ex}}$, the anisotropic energy $E_{\texttt{an}}$, the magnetostatic energy $E_{\texttt{m}}$, the Zeeman energy $ E_{\texttt{Zee}}$, and the interfacial Dzyaloshinskii-Moriya interaction (DMI) $E_{\texttt{DMI}}$. To effectively describe the steady-state motion of magnetic domains, Thiele proposed a collective coordinate method \cite{Thiele1973}. The key idea is to use parameter ${\bf R}(t)$ to represent domain's guiding center, such that ${\bf m}({\bf r},t)={\bf m}({\bf r}-{\bf R}(t))$. This approach treats the skyrmion, for instance, as a rigid body, and is applicable for describing its propagation and gyration, but not for the breathing motion, i.e., the periodic oscillation of the skyrmion radius. To simultaneously model the skyrmion gyration and breathing and their possible interactions, we generalize Thiele's approach by including the skyrmion radius degree of freedom
\begin{equation}\label{Eq2}
{\bf m}({\bf r},t)={\bf m}\Big (\frac{{\bf r}-{\bf R}(t)}{w(t)},\dot{w}(t)\Big),
\end{equation}
with $w$ and $\dot{w}$ being the skyrmion radius and its time derivative, respectively. Here, we did not take $\dot{\bf R}(t)$ as a collevtive cooridnate because the skyrmion gyration is much slower than its breathing [see Fig. \ref{Figure2}(c) below]. When considering the SkX, we adopt the subscript $j$ to label each skyrmion. 

We can write the total energy of SkX in a phenomenological way
\begin{equation}\label{Eq14}
\begin{aligned}
&W=\sum_{j}\frac{K_{b}}{2}w_{j}^{2}+\sum_{j}\frac{K_{g}}{2}\mathbf{U}_{j}^{2}+\sum_{\langle j\neq k\rangle}\frac{V_{jk}}{2},\\
\end{aligned}
\end{equation}
where $K_{b(g)}$ denotes the spring constant associated with the breathing (gyrating) motion of the $j$-th skyrmion, $w_j$ is the radius of the $j$-th skyrmion, $\mathbf{U}_{j}= \mathbf R_{j} - \mathbf R_{j}^{0}$ is the displacement of the $j-$th SK's guiding center $\mathbf{R}_{j}$ away from its equilibrium position $\mathbf{R}_{j}^{0}$ \cite{wang2018,li2021}, and
\begin{equation}\label{Eq15}
\begin{aligned}
V_{jk}=&\Gamma w_{j}w_{k}+k_{||}(w_{j}U_{k}^{||}+w_{k}U_{j}^{||})+k_{\bot}(w_{j}U_{k}^{\bot}+w_{k}U_{j}^{\bot})\\
&+\mathcal{I}_{\parallel}U_{j}^{\parallel}U_{k}^{\parallel}-\mathcal{I}_{\perp}U_{j}^{\perp}U_{k}^{\perp}+\mu_{\parallel}w_{j}w_{k}(U_{j}^{\parallel}+U_{k}^{\parallel})\\
&-\mu_{\perp}w_{j}w_{k}(U_{j}^{\perp}+U_{k}^{\perp})
\end{aligned}
\end{equation}is the interaction between two skyrmions labeled as $j$ and $k$, respectively. For simplicity, we only consider the nearest-neighbor coupling. Here, $\Gamma$ is the coupling coefficient of two breathing skyrmions, $U_{j}^{\parallel} \equiv \hat{e}_{jk}\cdot\textbf{U}_{j}$ is the projection of $\textbf{U}_{j}$ on the line connecting the two SKs with the unit vector $\hat{\mathbf{e}}_{jk}=(\mathbf{R}_{k}^{0}-\mathbf{R}_{j}^{0})/|\mathbf{R}_{k}^{0}-\mathbf{R}_{j}^{0}|$, and $U_{j}^{\perp}\equiv(\hat{z}\times \hat{e}_{jk})\cdot \mathbf{U}_{j}$ is the projection of displacement $\mathbf{U}_{j}$ perpendicular to $\hat{e}_{jk}$, $k_{\parallel}$ and $k_{\perp}$ describes the coupling parameter between the breathing and gyrating motion of two skyrmions, along the parallel and perpendicular direction, respectively, $\mathcal{I}_{\parallel}$ ($\mathcal{I}_{\perp}$) is the parallel (perpendicular) coupling constant of two gyrating skyrmions, and $\mu_{\parallel}$ ($\mu_{\perp}$) describes the nonlinear three-wave mixing that allows one breathing mode to be split into another breathing mode and a gyration mode. It is noted that one can safely neglect the $k_{\parallel}$ and $k_{\perp}$ terms because there exists a large gap between the skyrmion gyration and breathing, as shown in Fig. \ref{Figure2}(c) below. For the same reason, a breathing mode cannot be split into two gyration modes because of the constraint from the energy conservation. So, the $w_{j}U_{j}U_{k}$-type nonlinear interaction is not considered in our formalism. Equation \eqref{Eq15} then can be simplified as
\begin{equation}\label{Eq15S}
\begin{aligned}
V_{jk}=&\mathcal{I}_{\parallel}U_{j}^{\parallel}U_{k}^{\parallel}-\mathcal{I}_{\perp}U_{j}^{\perp}U_{k}^{\perp}+\mu_{\parallel}w_{j}w_{k}(U_{j}^{\parallel}+U_{k}^{\parallel})\\
&-\mu_{\perp}w_{j}w_{k}(U_{j}^{\perp}+U_{k}^{\perp})+\Gamma w_{j}w_{k}.
\end{aligned}
\end{equation}Based on Eqs. \eqref{Eq14} and \eqref{Eq15S}, we obtain the equations motion
\begin{equation}\label{Eq13}
\mathbf{G}\times {\dot{\mathbf{U}}}_{j}+\textbf{F}_{j}=0,
\end{equation}for skyrmion gyration, and
\begin{equation}\label{Eq13breathing}
\mathcal{M}\ddot{w_j}+K_{b}w_{j}+\sum_{k\in<j>}\Big[\Gamma  w_{k}+\mu_{\parallel}w_{k}(U_{j}^{\parallel}+U_{k}^{\parallel})-\mu_{\perp}w_{k}(U_{j}^{\perp}+U_{k}^{\perp})\Big ]=0,
\end{equation}
for skyrmion breathing, where $G= -4\pi Qd M_{s}$/$\gamma$ is the gyroscopic coefficient coefficient with $Q=\pm1$ the skyrmion charge, $d$ the thickness of the magnetic film, $\gamma$ the gyromagnetic ratio, and $M_{s}$ being the saturation magnetization, $\textbf{F}_{j}=-\partial {W}/ \partial \mathbf U_{j}$ is the driving force, and $\mathcal{M}$ is the skyrmion mass.

To obtain an analytical understanding, we consider an one-dimensional (1D) SkX, such that $\mathcal{I}_{\perp}=\mu_{\perp}=0$ and $\mathcal{I}_{\parallel}=\mathcal{I}$. In the linear region, we assume $\mu_{\parallel}=0$. Imposing ${\bf U}_j=(u_{j},v_{j})$ and $\psi_{j}=u_{j}+iv_{j}$, we have
\begin{subequations}
\begin{align}
 \dot{\psi}_{j}+ic_{1}\psi_{j}+ic_{2}(\psi_{j-1}+\psi_{j+1})=0, \label{eq:17a} \\
    \ddot{w}_{j}+c_{3}w_{j}+c_{4}(w_{j-1}+w_{j+1})=0, \label{eq:17b}
\end{align}
\end{subequations}
with parameters $c_{1}=K_g/G$, $c_{2}=\mathcal{I}/G$, $c_{3}=K_b/\mathcal{M}$, and $c_{4}=\Gamma/\mathcal{M}$. Next, we consider the plane-wave expansion of  $\psi_{j},w_j\propto$exp[$i(jka-\omega t)]$, where $k$ represents the wave vector, and $a$ is the lattice constant. From Eq. (\ref{eq:17a}), we derive the dispersion relation of the collective gyration motion
\begin{equation}\label{Eq18}
\omega_{g}=c_{1}+2c_{2}\cos(ka).
\end{equation}

From Eq. (\ref{eq:17b}), we obtain the following dispersion relation of the collective breathing motion
\begin{equation}\label{Eq19}
\omega_{b}=\sqrt{c_{3}+2c_{4}\cos(ka)}.
\end{equation}

To verify our theory, we implement full micromagnetic simulations below.
\begin{figure}[ptbh]\label{Figure1}
\begin{centering}
\includegraphics[width=0.5\textwidth]{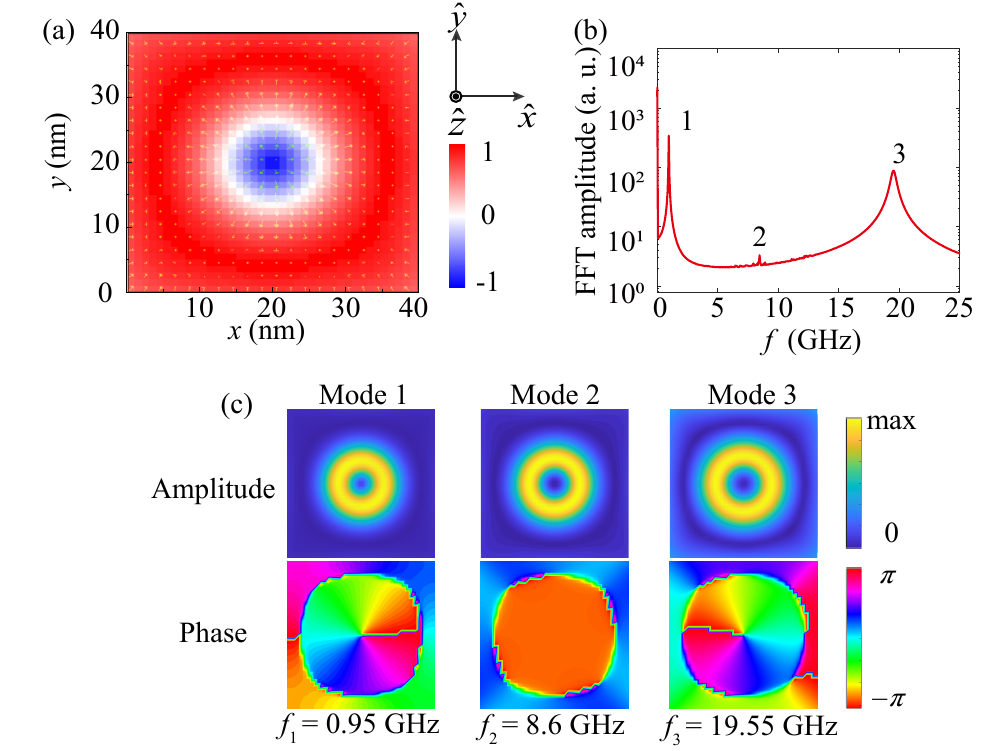}
\par\end{centering}
\caption{(a) Spatial configuration of $m_z$ of a single skyrmion in a square film with the sidelength 40 nm and thickness 1 nm. The blue (red) color stands for the spatial distribution of $m_{z}$. (b) Spectral power of a single skyrmion excited by an in-plane AC magnetic field. (c) Spatial maps of the $\delta m_{z}$ of frequency 0.95, 8.6, and 19.55 GHz, respectively. Top and bottom rows show the spectral amplitude and phase, respectively.}
\label{Figure1}
\end{figure}
\section{Micromagnetic simulations} \label{Micromagnetic simulations}
Micromagnetic simulations are performed by employing the open-source software $\texttt{MUMAX3}$ package \cite{Vansteenkiste2014}, which numerically solves the LLG equation \cite{Landau1935}. In the simulations, we adopt material parameters of Co/Pt \cite{sampaio2013nucleation}: the perpendicular anisotropy constant $K_{\texttt{u}}=0.6\times10^{6}$ J/m$^{3}$, the exchange constant $A_{\texttt{ex}}=1.5\times10^{-11} $ J/m, the saturation magnetization $M_{s}=5.8\times10^{5} $ A/m, the interfacial DMI strength $D=3$ mJ/m$^{2}$, and the damping constant $\alpha = 0.001$ if not stated otherwise. The cell size is set to be $1 \times1 \times 1$ nm$^{3}$.

To obtain the spin wave (SW) spectrum, we apply a sinc-function magnetic field $\mathbf{B}(t)=B_{0}\sin{[2\pi f_{0}(t-t_{0})]/ [2\pi f_{0}(t-t_{0})]}\hat{x}$ over the whole film, with the amplitude $B_{0} =5$ mT, $t_{0}=1$ ns and the cutoff frequency $f_{0} = 50$ GHz for 40 ns. The evolution of the magnetization is recorded per 5 ps and the power spectral is obtained by using fast Fourier transformation (FFT) \cite{Wanghao2007,BuessPRL2005}. The frequency resolution of the spectrum obtained is 0.025 GHz.

We first study the dynamic properties of a single skyrmion in a $40\times 40$ nm$^2$ film, as shown in Fig. \ref{Figure1}(a). We apply the in-plane AC field on the individual skyrmion, with their excitation spectra plotted in Fig. \ref{Figure1}(b). We observe three prominent resonant peaks at 0.95, 8.6 and 19.55 GHz, which are marked as 1, 2 and 3, respectively. The spectral information of those modes including their amplitudes and phases are shown in Fig. \ref{Figure1}(c). Modes 1 and 3 are counterclockwise and clockwise gyration modes, respectively, with the spectral phase winding 2$\pi$ around the center. For the mode 2, the spectral amplitude exhibits a radial symmetry, and the phase maintains as a constant. These features indicate the periodic expansion and compression of the skyrmion, which represents a skyrmion breathing.

\subsection{Collective excitations of 1D SkX}

\begin{figure}[ptbh]\label{Figure2}
\begin{centering}
\includegraphics[width=0.5\textwidth]{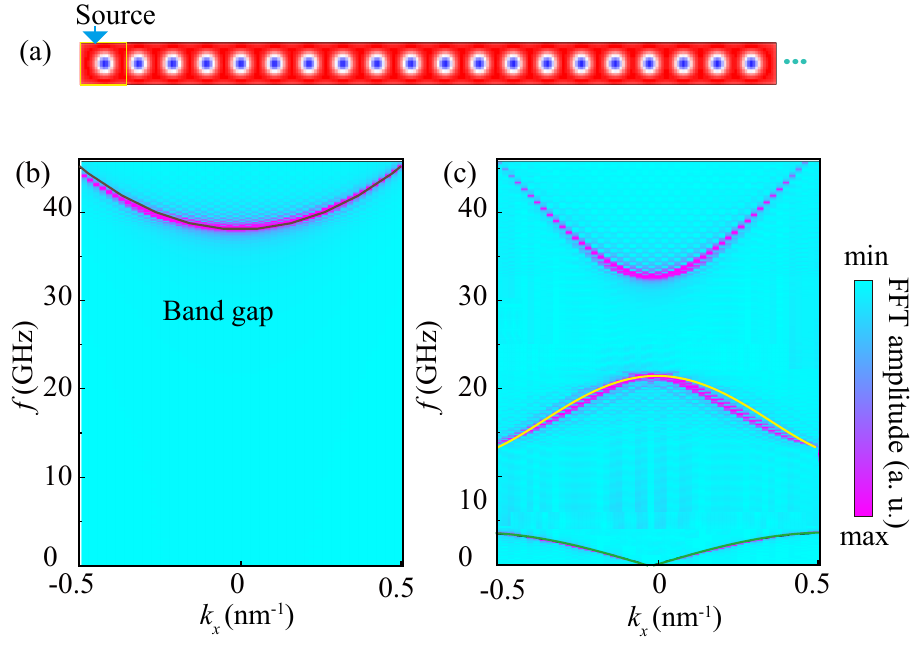}
\par\end{centering}
\caption{ (a) Schematic diagram of a 1D SkX. The yellow rectangle represents the position of the excitation source.  (b) Dispersion of the SW continuum in the FM state. The black curve represent the quadratic dispersion with the band gap $\omega_c$. (c) The dispersion relations of collective excitations in the SkX. The uppermost, middle, and bottom branches represent the SW continuum, skyrmion breathing, and skyrmion gyration bands, respectively. The green and yellow curves represent the analytical formulas (\ref{Eq18}) and (\ref{Eq19}), respectively.}
\label{Figure2}
\end{figure}

In this section, we extend the single skyrmion to 1D SkX. Figure \ref{Figure2}(a) shows the distribution of $m_{z}$ in an aray of skyrmions with a lattice constant $a=40$ nm and thickness $d=1$ nm. To study the dispersion relation, an in-plane sinc-function magnetic field with amplitude $B_{0}=5$ mT and cutoff frequency $f_{0}=30$ GHz has been applied locally to the left region with a width of 40 nm (as indicated by yellow rectangle). Figures \ref{Figure2}(b) and (c) show both the simulated and calculated dispersion of dynamic ferromagnetic (FM) state and 1D SkX. For the case of FM state, the lowest-frequency mode at $k=0$ is the FMR mode, whose frequency can be analytically calculated as $\omega_{c}/2\pi=\gamma(2K-\mu_{0}M^{2}_{s})/(2\pi M_{s})=37.5$ GHz. There is a continuous quadratic dispersion above 37.5 GHz, which reflects the standard behavior of the uniform magnetized magnon waveguide.

\begin{figure}[ptbh]\label{Figure3}
\begin{centering}
\includegraphics[width=0.5\textwidth]{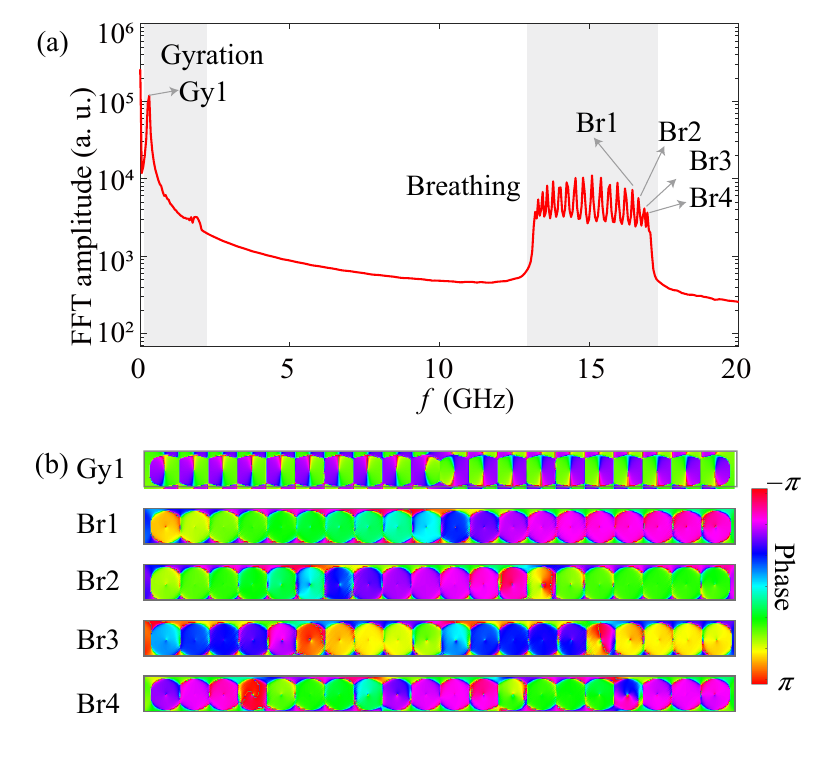}
\par\end{centering}
\caption{ (a) The excitation spectra for 1D SkX excited by an out-of-plane AC magnetic field. (b) The spatial profiles of the spectral phases for several representative resonant modes. 
\label{Figure3}}
\end{figure}

For SkX, we calculate the $f$-$k$ relation. As shown in Fig. \ref{Figure2}(c), the magnon dispersion possesses multiple magnonic bandgaps, which result from the scattering of magnons by periodically arranged skyrmions. This is similar to the results of the Kronig-Penney model for particles in a 1D lattice in quantum mechanics. We observe that the original FMR has shifted down to 34.5 GHz due to the presence of skyrmions. Interestingly, we find two new bands 0.1$\sim$4 GHz and 13.4$\sim$20.4 GHz, both of which are well below 34.5 GHz. The existence of these modes is contingent upon the presence of SkX. Without SkX, these patterns disppear. We thus envision that these branches stem from coupled skyrmion oscillations. We use formulas (\ref{Eq18}) and (\ref{Eq19}) to fit the dispersion curve of the two emerging bands obtained from micromagnetic simulations. The green and yellow curves in Fig. \ref{Figure2}(c) represent theoretical formulas $\omega_{g}=c_{1}+2c_{2}\cos(ka)$, and $\omega_{b}=\sqrt{c_{3}+2c_{4}\cos(ka)}$, respectively, which nicely match simulation results with fitting parameters $c_{1}=1.15$ GHz, $c_{2}=-0.5$ GHz, $c_3=294.9$ (GHz)$^2$ and $c_4=124$ (GHz)$^2$. In this regard, the 0.1$\sim$4 GHz band represents the skyrmion gyration band, while the 13.4$\sim$20.4 GHz band indicates the skyrmion breathing band.  It is conventionally believed that SWs below the FMR cannot propagate in thin ferremagnetic films. However, using the SkX as an effective medium, we achieve the propagation of both the breathing and gyration modes. Below, we provide more evidences to justify this point.

Figure \ref{Figure3}(a) shows the internal spectrum of 1D SkX by performing a FFT to each cell and averaging over the whole waveguide. We plot the spatial amplitude and phase distributions of the lowest peak and four highest peaks, as shown in Fig. \ref{Figure3}(b). We observe that the lowest frequency $f_{g}=0.3$ GHz is the collective gyration mode (Gy1) with the wavelength $\lambda=10\times40$ nm $=400$ nm and wavevector $k=0.0157$ nm$^{-1}$. For other higher frequency modes (Br1, Br2, Br3, and Br4) between 10$\sim$20 GHz, we identify them as collective breathing modes forming standing waves with different wavelengths. The interval between the collective breathing modes coincides with $f_{g}=0.3$ GHz. We thus conclude that the two low-lying bands observed below the FMR in Fig. \ref{Figure2}(c) represent the dispersion relations of collective skyrmion gyrating (0.1$\sim$4 GHz) and breathing (13.4$\sim$20.4 GHz) motions.

\begin{figure}[ptbh]\label{Figure4}
\begin{centering}
\includegraphics[width=0.49\textwidth]{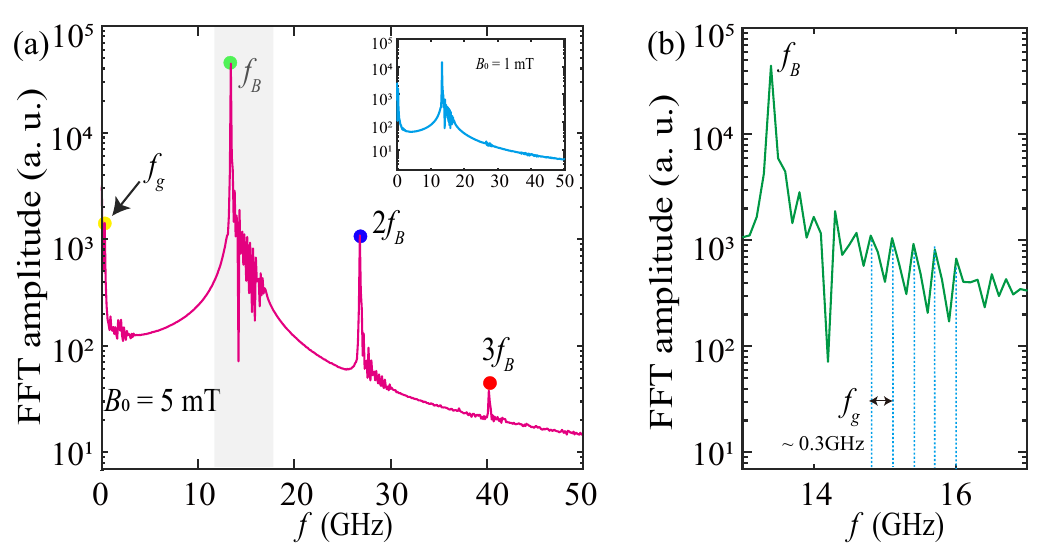}
\par\end{centering}
\caption{ (a) The system response to the microwave source with the driving amplitude $B_{0}=5$ mT and frequency of $f_{B}=13.4$ GHz. Inset: Spectrum with $B_{0}=1$ mT. (b) Magnified view of the gray spectrum region in (a).}
\label{Figure4}
\end{figure}

\subsection{MFC in 1D SkX}

As pointed out above, there exists a large gap between the gyration and breathing bands. A direct coupling between them is not possible, while an in-direct coupling is in principle allowed. One can imagine such a non-resonant process that one high-frequency breathing mode is split into another low-frequency breathing mode and a gyration mode, i.e.,
\begin{equation}\label{3magnon}
\omega_b\rightarrow\omega_{b'}+\omega_g,
\end{equation} or the other way around. The second  harmonic is also possible:
\begin{equation}\label{SHG}
\omega_b+\omega_b\rightarrow2\omega_b.
\end{equation} The chain-like process of these three-wave mixings finally generates the MFC. To demonstrate these processes, we excite the breathing mode of the 1D SkX by using a single-frequency microwave field $\mathbf{B}=B_{0}$sin$(2\pi f_{B}t)\hat{z}$ with $f_{B}=13.4$ GHz, in a narrow rectangular area of the film [yellow rectangle in Fig. \ref{Figure2}(a)]. Localized breathing oscillations will propagate through the system as a result of the skyrmion-skyrmion coupling. We make FFT of the time-dependent magnetization and observe significant nonlinear effects. When the amplitude of the microwave magnetic field is low ($B_{0}=1$ mT), the nonlinear effects are weak, and only the input frequency $f_{B}$ is observed [see the inset of Fig. \ref{Figure4}(a)]. As the field amplitude increases to 5 mT, higher-harmonics of the breathing mode emerge ($f_{l}^{B}=lf_{B}$, $l=1,2,3,...$ with different node numbers), as illustrated in Fig. \ref{Figure4}(a). The lowest peak $f_{g}=0.3$ GHz is the collective gyration mode. We have also observed dense peaks between the fundamental harmonic $f_{B}$ and the second harmonic $2f_{B}$, forming the comb-like SW spectrum. For enhanced observation of the low-frequency MFC, we have zoomed in on the gray segment of Fig. \ref{Figure4}(a), depicted in Fig. \ref{Figure4}(b).  \begin{figure*}[ptbh]\label{Figure5}
\begin{centering}
\includegraphics[width=1\textwidth]{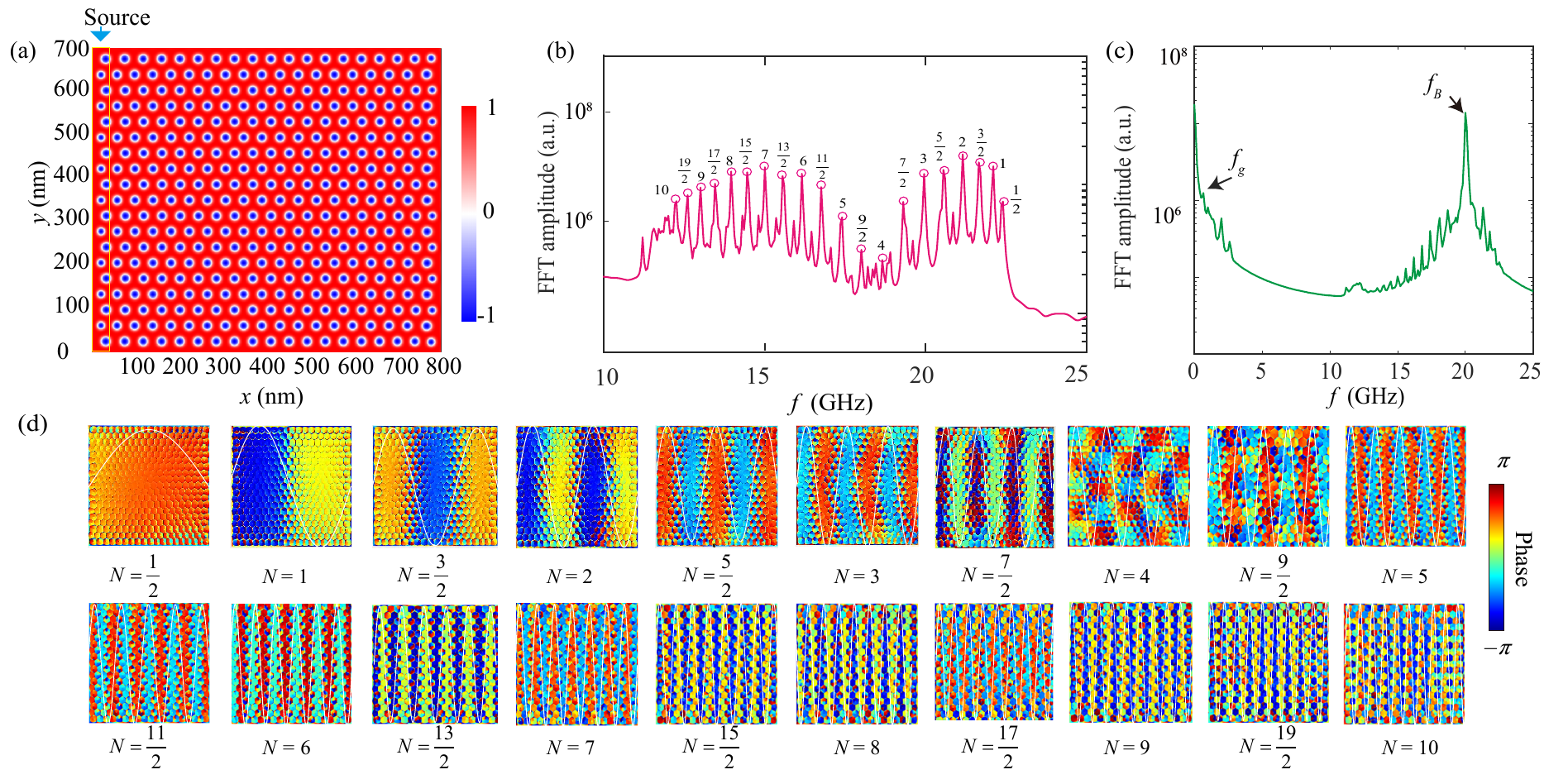}
\par\end{centering}
\caption{(a) The schematic diagrams of 2D SkX with the lattice constant $40$ nm and thickness $1$ nm. The yellow rectangle represents the area of the excitation source. (b) The excitation spectra by an in-plane sinc-function magnetic field. (c) The spectra driven by the harmonic field $B=B_{0}\sin(2\pi f_{B}t)$ along the $z$-axis with $B_{0} = 5$ mT and $f_{B}=20$ GHz. (d) The real-space profile of the spectral phase of 2D SkX. }
\label{Figure5}
\end{figure*}The frequency interval is equal to 0.3 GHz. These peaks correspond to standing wave patterns of the breathing modes, which can be well described by $f_{m}^{\texttt{MFC}}=f_{B}+mf_{g}$ with $m=0,1,2,3,...$. One does not observe the mode below $f_B$ because we chose the band-bottom frequency (13.4 GHz). We also note the MFC around the second-harmonic peak with the spacing being $f_g$ too.


Next, we analyze the origin of the MFC: The AC magnetic field first excites the collective skyrmion breathing in the 1D SkX, which forms a standing wave due to the constraint of the boundary. It is thus natural that the skyrmions at the nodes do not breathe. When the field amplitude $B_0$ is large enough, the gyration motion of the node-skyrmion can be excited due to the $\mu$-terms in the interaction \eqref{Eq15S}. The low-lying frequency comb then emerges because of the nonlinear mixing between the collective breathing and gyration modes in the SkX, a phenomenon that is absent for an individual skyrmion. It is noted that the minimum frequency of the gyration band is given by $c_{1}+2c_{2}=(K_g+2\mathcal{I})/G$. In the present setup, it is $0.15$ GHz, which sets a lower limit for the MFC spacing. One cn expect a close-to-gapless gyration band by further optimizing the SkX design, to obtain denser MFC with more teeth.


\subsection{MFC in 2D SkX}
It is straightforward to extend the above results to a 2D SkX. As a typical example, we consider a 2D SkX containing $19\times19$ skyrmions. To analyze the skyrmion breathing spectrum, we apply a microwave field $\mathbf{B}(t)=B_{0}$sinc$(2\pi f_{0}t)\hat{x}$ with $f_{0}=30$ GHz and $B_{0}=5$ mT locally covering the rectangle region of 2D SkX, as shown in Fig. \ref{Figure5}(a). The spectrum in Fig. \ref{Figure5}(b) reveals distinct peaks labeled by $N$ (where $N$ = $\frac{1}{2}, 1, \frac{3}{2}, 2, \frac{5}{2}, 3,...,10$) denoting the ratio between the side-length of the film and the wavelength of each breathing mode. To induce the MFC, we apply an oscillating magnetic field $\mathbf{B}(t)=B_{0}$sin$(2\pi f_{B}t)\hat{z}$ with $B_{0}=5$ mT and $f_{B}=20$ GHz (corresponding to $N=3$) over the narrow rectangular area depicted in Fig. \ref{Figure5}(a). We observe the emergence of frequency comb between 11.2 and 23.3 GHz, as shown in Fig. \ref{Figure5}(c). The spatial phase distributions of the $\delta m_{z}$ [shown in Fig. \ref{Figure5}(d)] in the whole SkX represent specific waves of different wavelengths. The breathing amplitudes of all peaks from 12.2 to 23 GHz are either symmetric or anti-symmetric about the lattice's center, with both ends being fully pinned. These features represent the interference between the forward-propagating  and backward-propagating breathing modes, which forms a standing wave. Similar to the 1D case, skyrmions at the nodes do not breathe but gyrate. We find that the peak interval is approximately $0.6$ GHz, the gyration frequency $f_g$ of the skrymion. This coincidence indicates a nonlinear interaction between gyration and breathing modes, which induces the MFC in the 2D SkX.

In the above discussions, we did not consider the thermal effect. The role of thermal fluctuations is twofold: First, it may lower the stability of the SkX; Second, the gyration mode is occupied at a finite temperature, even without the three-wave mixing, which may significantly lower the critical microwave power in producing the MFC \cite{Wang2022,Liang2024}. The disorder effect \cite{Reichhardt2022} is also ignored in the present work. We leave these issues for future study.
 
\section{Conclusion} \label{conclusion}
In summary, we theoretically studied the spectra of gyration and breathing modes, and their coupling in SkX by generalizing Thiele's collective-coordinate method. A large gap separating the high-energy breathing and low-energy gyration bands was identified, which excludes the resonant coupling between the two bands but allows their indirect coupling. We predicted the emergence of MFC far below the FMR. The origin of the low-lying MFC arises from three-wave mixing between the skrymion breathing and gyration at the standing-wave nodes both in 1D and 2D SkXs. Theoretical predictions were verified by full micromagnetic simulations with good agreement. Our results open the door for achieving SkX-based ultra-low energy and ultra-dense MFC that is indispensable for magnon precision measurements and information processing.

\begin{acknowledgments}
\section*{ACKNOWLEDGMENTS}
We thank Z. Wang and Z. Li for helpful discussions. This work was supported by the National Key R$\&$D Program China under Contract No. 2022YFA1402802 and the National Natural Science Foundation of China (NSFC) (Grants No. 12374103 and No. 12074057). X. J. Liu acknowledges financial support from the talent introduction program of Chengdu Normal University (No. YJRC 2021-14).
\end{acknowledgments}

\end{document}